\documentclass[11pt,a4paper]{article}

\usepackage{a4wide}
\usepackage[fleqn]{amsmath}
\usepackage{hyperref}
\usepackage[papersize={180mm,240mm}]{geometry}
\usepackage{float}
\usepackage[numbers]{natbib}
\usepackage{multirow}

\usepackage{fourier} 
\usepackage[scaled=0.875]{helvet} 

\RequirePackage{amsmath,amssymb,amsxtra,amsthm}

\RequirePackage{mathrsfs}

\RequirePackage{setspace}
\RequirePackage{array}
\RequirePackage{booktabs}

\RequirePackage{braket}

\RequirePackage[pdftex,final]{graphicx}
\graphicspath{{Plots/}}

\RequirePackage{colortbl}
\RequirePackage{xcolor}
\colorlet{titlerowcolor}{gray!15}

\newcommand{\be}{\begin{equation}}
\newcommand{\ee}{\end{equation}}
\newcommand{\bea}{\begin{eqnarray}}
\newcommand{\eea}{\end{eqnarray}}

\usepackage{authblk}
\title{Non-perturbative Topological String Partition Function on Twisted Affine Line Bundle over $\mathbb{C}\times T^2$ }

\author[1]{Ignatios Antoniadis\thanks{antoniad@lpthe.jussieu.fr}}
\author[2]{Marine Samsonyan\thanks{m.samsonyan@yerphi.am}}
\affil[1]{Institute for Advanced Study, 1 Einstein Drive, Princeton, New Jersey 08540,  USA\\

  LPTHE, Sorbonne University, CNRS, 4 Place Jussieu, 75005  Paris, France}
\affil[2]{A. I. Alikhanyan National Science Laboratory, 2 Alikhanian Br., 0036\\

Yerevan State University, 1 Alek Manukyan St, 0025 Yerevan, Armenia}

\begin{document}
\date{}
\maketitle

\begin{abstract}
 Using instanton partition function for five dimensional $U(1)$ 
 gauge theory with eight supercharges and a single adjoint massive hypermultiplet on the $\Omega$ background, we give explicit expression for
 non-perturbative corrections to the topological string theory in the holomorphic limit.  It was argued that in this case the theory is compactified on the twisted affine line bundle over $\mathbb{C}\times T^2$. We perform calculations in two ways. First we modify the integration contour by adding poles responsible for non-perturbative physics in accordance with a recent proposal. Then,  we compute the genus zero Gopakumar-Vafa invariants for our case and evaluate the non-perturbative corrections to the partition function. We check that both calculations give the same result.
\end{abstract}

\newpage
\tableofcontents
\section{Introduction}

It is known that the perturbative topological string free energy $F^p$ is given as an asymptotic series in the topological string coupling $\lambda$, $F^p=\sum_{g=0}^\infty F_g \lambda^{2g-2}$. It has been shown in \cite{Antoniadis:1993ze} 
that certain type II string amplitudes at genus $g$ reproduce the topological string partition function $F_g$  \cite{Bershadsky:1993ta} and generate higher-derivative F-terms in the four-dimensional effective action of the form 
$F_g W^{2g}$, where $W$ is the ${\cal N}=2$ chiral Weyl superfield whose lowest component is the (self-dual) graviphoton field strength. The topological string coupling is identified with the graviphoton background.

In \cite{Gopakumar:1998ii, Gopakumar:1998jq}, it has been shown that $F^p$ encodes the BPS spectrum of M2-branes wrapped on two-cycles of a Calabi-Yau (CY) threefold in M-theory compactifications. In particular, this 
framework naturally engineers ${\cal N}=2^*$ gauge theories, e.g. \cite{Iqbal:2007ii, Haghighat:2013gba} in which the gauge theory degrees of freedom originate from M2-branes wrapping non-trivial holomorphic two-cycles in the CY geometry.

More recently, \cite{Hattab:2024ewk} proposed a method to extract the non-perturbative (in $\lambda$) contributions to the topological string free energy from the perturbative expansion \cite{Gopakumar:1998jq} by carefully 
treating additional poles into the Schwinger proper-time integral.

On the other hand, in \cite{Alim:2024dyi} a closed-form expression for the non-perturbative topological string partition function on CY threefold was derived in the holomorphic limit. It was 
shown that the result admits a factorized form in which each factor is given by a shifted resolved conifold partition function raised to an appropriate power. Remarkably, the non-perturbative corrections involve only genus-zero 
Gopakumar-Vafa (GV) invariants.

In what follows, we compute the non-perturbative partition function for topological string theory on a specific CY manifold. In the field-theory limit, this geometry corresponds to five-dimensional ($5D$) $U(1)$ ${\cal N}=2^*$ gauge theory placed on the 
$\Omega$-background \cite{Nekrasov:2003rj}, \cite{Nekrasov:2002qd}  with a single deformation parameter. The string theory realisation of the perturbative part of this theory with two deformation parameters has been given in \cite{Angelantonj:2017qeh}, \cite{Samsonyan:2017xdi}.  We use the instanton partition function of the $5D$ $U(1)$ ${\cal N}=2^*$  theory computed via localization \cite{Poghossian:2008ge} and expressed in the topological vertex formalism parameters \cite{Aganagic:2003db}.
Starting from this expression, and following the prescription of \cite{Hattab:2024ewk}, we determine the additional poles required to reproduce the full non-perturbative topological string partition function and compute the associated residues explicitly. We then compare our result with the universal form proposed in \cite{Alim:2024dyi}, in which the non-perturbative partition function is constructed solely from resolved conifold GV invariants. For the geometry under consideration, we determine the relevant invariants and find complete agreement between the two approaches.

\section{Partition functions}
\label{5dN=2*}

\begin{figure}[h]
    \centering
    \includegraphics[width=0.4\textwidth]{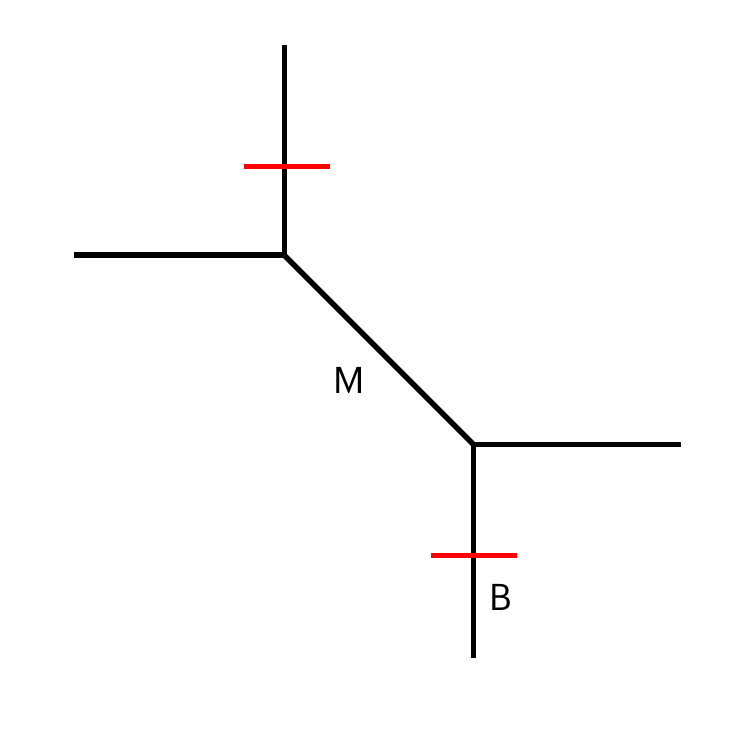}
    \caption{Toric diagram for ${\cal N}=2^*$ $U(1)$ theory. We use the cycle basis $E=B+M$ and $M$. \label{U1tor}}
\end{figure}

The instanton partition function of $\Omega$-deformed $5D$ Abelian gauge theory with an extra adjoint hypermultiplet reads (Eq. (2.10) of \cite{Poghossian:2008ge})
\begin{eqnarray}
Z_{Adj}^{U(1), 5D}=\exp\left(\sum_{n=1}^\infty\frac{{\bf Q}^n\left(1-(T_m T_1)^n\right) \left(1-(T_m T_2)^n\right)}{n \left(1-T_1^n\right) \left(1-T_2^n\right) \left(1-({\bf Q}T_m )^n\right)}\right)\label{PS}\,.
\end{eqnarray}
Hence the free energy in this case has the following form
\begin{eqnarray}
F=\log Z=\sum_{n=1}^\infty\frac{{\bf Q}^n\left(1-(T_m T_1)^n\right) \left(1-(T_m T_2)^n\right)}{n \left(1-T_1^n\right) \left(1-T_2^n\right) \left(1-({\bf Q} T_m )^n\right)}\label{Fnp}\, ,
\end{eqnarray}
where ${\bf Q}$ is the instanton counting parameter (or one of the exponentiated K\"ahler parameters), see the Appendix for different notations in the literature.  $T_1=e^{i\epsilon_1 R}$, $T_2=e^{i\epsilon_2 R}$, $T_m=e^{-m R}$, where $\epsilon_{1,2}$ are deformation parameters of the $\Omega$ background, $m$ is the mass parameter of the massive hypermultiplet (the other K\"ahler parameter of the geometry), and $R$ is the radius of the fifth compact dimension. Note that the four dimensional theory can be obtained by tuning the parameters appropriately, and taking the $R\to 0$ limit, see the Appendix. 
To make contact with the topological vertex notation, we need to replace $
T_1=t, T_2=q^{-1}, T_m=Q_m$.
In these notations, \eqref{Fnp} can be rewritten in the following form
\begin{eqnarray}
F&=&\sum_{n=1}^\infty \frac{{\bf Q}^n \left(1-Q_m^n t^n\right) \left(1-Q_m^n q^{-n}\right)}{n \left( 1-t^n\right)\left(1-q^{-n}\right)  \left(1-({\bf Q} Q_m)^n\right)}\nonumber\\
&=&-\sum_{n=1}^\infty \frac{\left(q t^{-1}\right)^{n/2} {\bf Q}^n \left(1-Q_m^n t^n\right) \left(1-Q_m^n q^{-n}\right)}{n \left(1-({\bf Q} Q_m)^n\right)\left( t^{n/2}-t^{-n/2}\right)\left(q^{n/2}-q^{-n/2}\right)}\, .\label{Finstop}
\end{eqnarray}
The case of one parameter deformation $\epsilon_1=-\epsilon_2$, or $T_1 T_2=1$, corresponds to $t=q$, and in this case one finds
\begin{eqnarray}
F&=&-\sum_{n=1}^\infty \frac{{\bf Q}^n \left( 1-Q_m^n q^n\right)\left(1-Q_m^n q^{-n}\right)}{n \left(1-({\bf Q} Q_m)^n\right)\left(q^{n/2}-q^{-n/2}\right)^2}\label{PSnph}\\
&=&-\sum_{n=1}^\infty \frac{\left({\bf Q} Q_m\right)^n}{n \left(1-\left({\bf Q} Q_m\right)^n\right)}\frac{Q_m^n+Q_m^{-n}-(q^n+q^{-n})}{ \left(q^{n/2}-q^{-n/2}\right)^2}\label{PSnph1}\\
&=&-\sum_{n,\ell =1}^\infty \frac{Q^{\ell n}Q_m^n+ Q^{\ell n} Q_m^{-n}-Q^{\ell n} (q^n+q^{-n})}{ n \left(q^{n/2}-q^{-n/2}\right)^2}\label{PSnph2}\,, 
\end{eqnarray}
where we introduced $Q={\bf Q} Q_m$ notation. Later, we will use the fact that ${\bf Q} Q_m$ corresponds to the K\"ahler parameter associated with a 2-cycle $E$ and $Q_m$ is the K\"ahler parameter associated with a 2-cycle $M$ \cite{Haghighat:2013gba, Huang:2013yta} Figure \ref{U1tor}.

In the following we want to use the procedure of \cite{Hattab:2024ewk} and find the extra poles. Calculating the residues on these poles will give the non-perturbative (in $\lambda$) corrections to the topological free energy. For this purpose, let us make the following notations $Q=e^{-\tilde{t}}$, $T_1=T_2^{-1}=q=e^{i\lambda}$, $Q_m=e^{-mR}$, which brings \eqref{PSnph1} to the form
\begin{eqnarray}
F&=&-\sum_{n=1}^\infty\frac{e^{-\tilde{t}n}}{n (1-e^{-\tilde{t}n})} \frac{e^{-mRn}+e^{mRn}-(e^{i\lambda n}+e^{-i\lambda n})}{(e^{i\lambda n/2}-e^{-i\lambda n/2})^2}\nonumber\\
&=&2\sum_{n=1}^\infty\frac{e^{-\tilde{t}n}}{n (1-e^{-\tilde{t}n})} \frac{\cosh(mRn)-\cos(\lambda n)}{ \left(2 \sin(\lambda n/2)\right)^2}\label{2cos}\,.
\end{eqnarray}
Note that the sum converges because $\tilde{t}$ is bounded from below by $mR$ (by the definition of $Q={\bf Q} Q_m$). The sum can now be rewritten as a contour integral 
\begin{eqnarray}
F=2 \oint \frac{ds}{s} \frac{1}{1-e^{-2\pi i s}} \frac{1}{e^{\tilde{t} s}-1 }\frac{\cosh(mRs)-\cos(\lambda s)}{\left( 2 \sin(s\lambda/2)\right)^2}=\oint f(s) ds \, . \label{fsint}
\end{eqnarray}

The perturbative (no-instanton) free energy for ${\cal N}=2^*$ $U(1)$ theory was realized and computed within a string theory setup in  \cite{Angelantonj:2017qeh}. In particular, in the case of one deformation parameter $\epsilon_1=-\epsilon_2=\hbar$, it can be written in the following form:  
\begin{eqnarray}
{\cal{F}}_{\text{pert.}}=-2 \int_{-\infty}^{+\infty}\frac{dt}{t} \frac{1}{1-e^{-t}}\frac{\cos(mR t)-1}{\sinh^2 (2R\hbar t)}\label{Fpert}\, .
\end{eqnarray}
Indeed, from eqs.(3.11), (3.13) in \cite{Angelantonj:2017qeh}, one obtains:
\begin{eqnarray}
{\cal{F}}_{\text{pert.}}=-\int_0^\infty \frac{dt}{t}\frac{1}{\sinh^2(2R\hbar t)}\left[-2\left(\sum_{l=0}^\infty e^{-t s}+\sum_{l=1}^\infty e^{-tl}\right)\right.\nonumber\\
\left.+\sum_{l=0}^\infty e^{-t (l-i mR)}+\sum_{l=1}^\infty e^{-t (l+i mR)}\right.\nonumber\\
\left.+\sum_{l=0}^\infty e^{-t (l+i mR)}+\sum_{l=1}^\infty e^{-t (l-i mR)}\right]\,.
\end{eqnarray}
One can now extend the sums from $l=1$ to $l=0$ and combine $e^{\pm imR t}$ into a cosine:
\begin{eqnarray}
{\cal{F}}_{\text{pert.}}=-\int_0^\infty \frac{dt}{t}\frac{1}{\sinh^2(2R\hbar t)}\left[-2\left(2\sum_{l=0}^\infty e^{-tl}-1\right)+2\cos(mRt)\left(2\sum_{l=0}^\infty e^{-tl}-1\right)\right]\, .
\end{eqnarray}
Summing over $l$ one then gets
\begin{eqnarray}
{\cal{F}}_{\text{pert.}}=2\left[-\int_0^\infty \frac{dt}{t}\frac{1}{\sinh^2(2R\hbar t)}\frac{\cos(mRt)-1}{1-e^{-t}}\right.\nonumber\\
+\left.\int_0^\infty \frac{dt}{t}\frac{1}{\sinh^2(2R\hbar t)}\frac{\cos(mRt)-1}{1-e^t}\right]\,.
\end{eqnarray}
Finally, in the second integral by changing $t\to-t$, we get \eqref{Fpert}.

One observes that the integrand \eqref{Fpert} can be obtained from \eqref{fsint} after writing $\cos(s\lambda)=1-2\sin^2(s\lambda/2)$ in \eqref{fsint} where the second term becomes $\lambda$- and $m$-independent and can be ignored for the rest of the argument. If one now deforms the contour around the positive semi-axis as an integral along the whole imaginary axis $2\pi i s=t$, making the identification $\lambda=8\pi R\hbar$, the two expressions match, up to the $\tilde{t}$-dependent factor in \eqref{fsint} containing the instanton counting parameter.  Formally, this can be done by taking the limit $e^{\tilde{t}} \to 0$, corresponding to $g_{YM}^2\to 0$ from the negative direction since $\tilde{t}\sim 1/g_{YM}^2$. Another way to obtain the perturbative result is by extrapolation of the instanton sum: expanding $\frac{1}{e^{\tilde{t} s}-1}=\sum_{n=1}^\infty e^{-\tilde{t} sn}$ in \eqref{fsint} and then extend the sum to $\sum_{n=0}^\infty e^{-\tilde{t} sn}$, where the $n=0$ term corresponds to \eqref{Fpert} upon the contour deformation described above.\footnote{An alternative method was proposed in~\cite{Hattab:2024yol} but it is not obvious how to apply it in the example under consideration.}
 
As mentioned above, the integrand in \eqref{fsint} is designed such that the residues at $s=n\in N$ exactly coincide with the summand in \eqref{2cos}. Alternatively, $f(s)$ in \eqref{fsint} can be obtained from eq.(22) in  \cite{Hattab:2024ewk} by inserting the appropriate GV multiplicities for our CY manifold (see later).  The contour in \eqref{fsint} should be chosen such that to surround the positive real axis without zero. It will include the poles at $s=n\in N$ and $s=\frac{2\pi k}{\lambda}$. The residues at $s=n\in N$ poles give the perturbative contribution. The non-perturbative corrections to the topological free energy can be found by calculating the residues at   $s=\frac{2\pi k}{\lambda}$. These residues are
\begin{eqnarray}
\left.\text{Res} f(s)\right\vert_{s=\frac{2\pi k}{\lambda}}=\left.\frac{d}{ds}\left( \frac{2(\cosh(smR)-\cos(s\lambda))}{s(1-e^{-2\pi i s})(e^{\tilde{t} s}-1)\lambda^2}\right)\right\vert_{s=\frac{2\pi k}{\lambda}}\, .
\end{eqnarray}
By defining the following sum
\begin{eqnarray}
\tilde{F}^{0}(\lambda,\tilde{t},u)=\sum_{k=1}^\infty\frac{i e^{-k\tilde{t}}\sinh^2(k\pi mR)}{2\pi^2 k^2 \sin(k\pi\lambda)(1-e^{-k(\tilde{t}+i\pi\lambda)})}\label{N=2starNS}\,\, ,
\end{eqnarray}
one observes that for our case the non-perturbative contributions to the free energy of the topological strings  will be given by
\begin{eqnarray}
\text{Res}\left. f(s)\right\vert_{s=\frac{2\pi k}{\lambda}}=\frac{d}{d\lambda}\left[\lambda \tilde{F}^0\left(\frac{2\pi}{\lambda}, \frac{2\pi}{\lambda}(\tilde{t}-i\pi), \frac{mR}{\lambda}\right)\right]\,.\label{111}
\end{eqnarray}

On the other hand, it is known that using $M2-M5$ branes setup one gets a $5D$ field theory with eight supercharges (in particular ${\cal N}=2^*$, which is our main interest) \cite{Iqbal:2007ii}, \cite{Haghighat:2013gba}.  $M2$ branes can wrap on various 2-cycles. Quantization of the moduli space of these branes gives rise to particles in the  $5D$ field theory, which carry $SU(2)_L\times SU(2)_R$ quantum numbers. In $5D$, $SO(4)=SU(2)_L\times SU(2)_R$ is the little group of massive particles. The contribution to the $F$-terms in the effective action from all particles is given by summing over the momentum, the holomorphic curves and the left-right spin content and takes the following form
\begin{eqnarray}
F=\sum_{\beta\in H^2(X,{\mathbb Z})} \sum_{n=1}^\infty \sum_{J_L} (-1)^{2 J_L} {\cal N}_\beta^{J_L} e^{-n T_\beta} \frac{q^{-2 J_L n}+....+q^{2 J_L n}}{n (q^{n/2}-q^{-n/2})^2}\label{tops}\, ,
\end{eqnarray}
where $H^2(X,{\mathbb Z})$ describes the $CY$ manifold and
\begin{eqnarray}
{\cal N}_\beta^{\,\, J_L}=\sum_{J_R}{\cal N}_\beta^{(J_L\, , J_R)} (-1)^{2 J_R} (2 J_R+1)\label{particles}
\end{eqnarray}
is the number of BPS states coming from an $M2$-brane wrapped around the holomorphic curve $\beta$ and the spin content under $SU(2)_L\times SU(2)_R$ is given by $(j_L, j_R)$.  The mass of the $M2$-brane wrapping a curve class $\beta\in H^2(X,{\mathbb Z})$ is given by $T_\beta$ and $q=e^{i \lambda}$. 
For the $5D$ ${\cal N}=2^*$ $U(1)$ theory the relevant  non compact CY threefold can be described as an affine line bundle over $\mathbb{C}\times T^2$ 
 with a nontrivial twist: the holonomy of the fiber around one 1-cycle of the torus is trivial, while transport around the other 1-cycle acts by a translation of the fiber coordinate by the adjoint mass parameter. Equivalently, parallel transport along this cycle induces a shift of the fiber by the mass $m$ (or its exponentiated version $Q_m$), which geometrically implements the ${\cal N}=2^*$ mass deformation.
 
 On the other hand, the topological string free energy can be written in the GV form as described in  \cite{Alim:2024dyi}
\begin{eqnarray}
F_{\text{GV}}(\lambda, t)=\sum_{\beta>0} \sum_{g\ge 0}[GV]_{\beta, g} \sum_{n\ge 1}\frac{1}{n \left(2\sin\left(\frac{n\lambda}{2}\right)\right)^2} (Q^{\beta})^n \left(2\sin\left(\frac{n\lambda}{2}\right)\right)^{2g}\,.\label{GVform}
\end{eqnarray}
It is possible to find a relation between $[GV]_{\beta, g}$ and ${\cal N}_\beta^{J_L}$ by comparing \eqref{tops} and \eqref{GVform} . We find 
\begin{eqnarray}
\sum_{g\ge 0}(-1)^{-g+1} [GV]_{\beta, g}\left( q^{1/2}-q^{-1/2}\right)^{2g}=(-1)^{2 J_L} {\cal N}_\beta^{J_L}\sum_{\ell =-J_L}^{J_L} q^{2\ell}\,.\label{GVN}
\end{eqnarray}
Indeed, this formula allows one to find $[GV]_{\beta, g}$ in our case. Remind that as independent cycles we use $E=B+M$ and $M$, see Figure \ref{U1tor}. Comparing \eqref{tops} with \eqref{PSnph2} , we see that the first term in the numerator of \eqref{PSnph2} corresponds to the cycle $\ell E+M$ with left spin $j_L=0$ (no $q$ dependence) and  ${\cal N}^0_{\ell E+M}=1$, the second term corresponds to $\ell E-M$ with left spin $j_L=0$ and ${\cal N}^0_{\ell E-M}=1$. The third term is the $\ell E$ cycle with the number of particles ${\cal N}^{1/2}_{\ell E}=1$ and left spin $j_L=1/2$ (the character is $q^n+q^{-n}$).
Hence, we have
\begin{eqnarray}
{\cal N}^0_{\ell E+M}={\cal N}^0_{\ell E-M}={\cal N}^{1/2}_{\ell E}=1\,.
\end{eqnarray}
In \cite{Alim:2024dyi}, it has been shown that  only genus zero GV invariants are needed to calculate the non-perturbative partition function for topological strings. We can use the relation \eqref{GVN} to find the GV invariants for $g=0$ in our case.  For ${\cal N}^0_{\ell E+M}=1$, we have $-[GV]_{\ell E+M,0}={\cal N}_{\ell E+M}^0=1$. For ${\cal N}^0_{\ell E-M}=1$, the equation gives\linebreak  $-[GV]_{\ell E-M,0}={\cal N}_{\ell E-M}^0=1$. Finally for  ${\cal N}_{\ell E}^{1/2}=1$, one gets \linebreak $-[GV]_{\ell E, 0}+[GV]_{\ell E,1} (q^{1/2}-q^{-1/2})^2=-{\cal N}_{\ell E}^{1/2}(q+q^{-1})$, which gives $[GV]_{\ell E,1}=-1$ and $[GV]_{\ell E, 0}=2$. Thus, all genus zero GV invariants that we need are
\begin{eqnarray}
[GV]_{\ell E+M, 0}=-1, \quad [GV]_{\ell E-M, 0}=-1, \quad [GV]_{\ell E, 0}=2\, .
\end{eqnarray}

As proven in \cite{Alim:2024dyi}, the formula to calculate the non-perturbative partition function for topological strings via only the  genus zero GV invariants is the following
\begin{eqnarray}
F_{np,top}(\lambda, \tilde{t})=-\frac{1}{2\pi i}\partial_\lambda\left(\lambda F^0 \left(\frac{2\pi}{\lambda},\left(\frac{t^\beta-i \pi}{\lambda}\right) 2\pi\right)\right)\, ,\label{invF}
\end{eqnarray}
where
\begin{eqnarray}
F^0=\sum_\beta [GV]_{\beta, 0} F(\lambda,t^\beta)=\frac{1}{2\pi}\sum_{\beta, n>0} \frac{[GV]_{\beta ,0}}{2 n^2}\frac{Q^{n\beta}}{\sin(\pi n\lambda)}\,  \label{F0res}
\end{eqnarray}
and $Q^{\beta}=e^{-t^\beta}$ with $t_\beta$  associated to two dimensional cycles. $F(\lambda,t^{\beta})$ is given by
\begin{eqnarray}
F(\lambda,t^{\beta})=\frac{1}{2\pi}\sum_{n>0}\frac{e^{-n t_\beta}}{2n^2\sin(n\pi\lambda)}\, .
\end{eqnarray}
 We found, that in our case the possible cycles and the corresponding K\"ahler parameters are 
\begin{eqnarray}
&&\beta=(\ell E+M, \ell E-M, \ell E) \label{cycles}\\
&& t_{\ell E+M}=\ell \tilde{t} +mR \\ 
&&t_{\ell E-M}=\ell \tilde{t} -mR \\
&& t_{\ell E}=\ell \tilde{t}\, .
 \end{eqnarray}
 Hence, for $Q^{\beta}$ we have
$Q^{\ell E+M}=e^{-(\ell \tilde{t}+mR)}$, $Q^{\ell E-M}=e^{-(\ell \tilde{t}-mR)}$ and $Q^{\ell E}=e^{-\ell \tilde{t}}$.
Then in this case $F^0\left(\frac{2\pi}{\lambda},\left(\frac{t^\beta-i \pi}{\lambda}\right) 2\pi\right)$ becomes
\begin{eqnarray}
F^0\left(\frac{2\pi}{\lambda}, \frac{2\pi (t^\beta-i\pi) }{\lambda}\right)=&&\hspace{-0.5cm}\sum_{\ell>0}\left[-F\left(\frac{2\pi}{\lambda},\frac{2\pi (\ell \tilde{t}+ mR -i\pi)}{\lambda}\right)\right.\nonumber\\
&&\left.\quad-F\left(\frac{2\pi}{\lambda},\frac{2\pi (\ell\tilde{t}-mR-i\pi) }{\lambda}\right)\right.\nonumber\\
&&\left.\quad+2 F\left(\frac{2\pi}{\lambda},\frac{2\pi (\ell \tilde{t}-i\pi) }{\lambda}\right)\right]\,.
\end{eqnarray}
It is now easy to compare this with \eqref{111}. One finds that they match
\begin{eqnarray}
F^0\left(\frac{2\pi}{\lambda}, \frac{2\pi (t^\beta-i\pi)}{\lambda}\right)=  2\pi i\tilde{F}^0\left(\frac{2\pi}{\lambda}, \frac{2\pi}{\lambda}(\tilde{t}-i\pi), \frac{mR}{\lambda}\right)\,.
\end{eqnarray}
Thus, we find exact agreement between the two approaches for the specific geometry under consideration.

\section{Conclusions}

The analysis carried out in this work focuses on a specific non-compact CY geometry which, in the appropriate field-theory limit, engineers the $5D$ $U(1)$  ${\cal N}=2^*$ gauge theory placed on an $\Omega$-background with a single deformation parameter. This background corresponds to turning on one equivariant parameter in the localization computation, and the resulting instanton partition function is known explicitly from both the Nekrasov localization approach and its reformulation in the language of the topological vertex.
Starting from this exact expression for the instanton partition function, we apply the prescription developed by Hattab and Palti to extract the full non-perturbative topological string partition function. The key step is to identify and incorporate the additional poles in the Schwinger proper-time representation that are not captured by the naive perturbative expansion. These extra poles encode the non-perturbative sectors in the topological string coupling, and their residues are computed explicitly.
Once these contributions are included, we obtain the complete non-perturbative partition function for this geometry. We then compare our result with the closed, factorized expression proposed by Alim, in which the non-perturbative partition function for any CY threefold is constructed entirely in terms of shifted resolved-conifold building blocks weighted by genus-zero GV invariants. For our geometry, we determine the relevant GV invariants and find perfect agreement between the two constructions. 

\section*{Acknowledgements}
 The research was supported by the Higher Education and Science Committee of MESCS RA ( Research project $\#$ 24RL-1C036). The authors are grateful to C. Angelantonj and H. Palti for valuable discussions.

\section*{Appendix}

For the reader convenience in the table below we present different notations, namely localization techniques as in \cite{Poghossian:2008ge}, topological string as in \cite{Hattab:2024ewk}, $5D$ Nekrasov partition function notation as in \cite{Nekrasov:2003rj} and refined topological vertex as in \cite{Iqbal:2007ii}.
\pagebreak

\begin{table}[htbp]
\centering
\caption{Notations\label{tab5}}
\label{tab5}
\begin{tabular}{cccc}
\hline
\textbf{Localization} &
\textbf{Topological Strings} &
\textbf{$5D$ Nekrasov} &
\textbf{Topological Vertex} \\
\hline
$T_1$ & $e^{i\lambda}$ & $e^{i\epsilon_1 R}$ & $t$ \\
$T_2$ & $e^{-i\lambda}$ & $e^{i\epsilon_2 R}$ & $q^{-1}$ \\
$T_m$ & \multirow{2}{*}{$e^{-\tilde{t}}$} & $e^{-m R}$ & $Q_m$ \\
$\mathbf{q}$ &  & $\Lambda$ & $\mathbf{Q}$ \\
\hline
\end{tabular}
\end{table}
When $m\rightarrow \infty$, $e^{-m}\rightarrow 0$ the hypermultiplet decouples and one recovers ${\cal N}=2$ theory. In the limit $m\rightarrow 0$ the ${\cal N}=4$ theory is obtained. 
$\Lambda$ is the gauge theory instanton counting parameter. In the case of pure gauge theory $\Lambda_{5D}\sim R^2 \Lambda_{4D}=R^2 e^{2\pi i \tau_{4D}}$ .
For the $5D$ theory with a single hypermultiplet under consideration  the instanton counting parameter $\Lambda_{5D}\sim \Lambda_{4D}$.
In this case, in going from $5D$ to $4D$ the factors $R^2$ emerge both in numerator (hypermultiplet part)
 and in denominator (vector multiplet part) in \eqref{PS} and cancel out. Hence, no further renormaliztaion of gauge coupling is needed.
 
 In the $4D$ limit $R\to 0$ and from \eqref{Fnp} one finds
\begin{eqnarray}
F=-\sum_{n=1}^\infty\frac{{\bf{Q}}^n}{n}\frac{(m-i \epsilon_1)(m-i \epsilon_2)}{\epsilon_1 \epsilon_2 (1-{\bf{Q}})^n}\,. 
\end{eqnarray}
The $m\to 0$ limit from the above formula gives 
\begin{eqnarray}
F=\sum_{n=1}^\infty \frac{{\bf{Q}}^n}{n(1-{\bf{Q}}^n)}=\sum_{n,k=1}^\infty\frac{({\bf{Q}}^k)^n}{n}=\sum_{k=1}^\infty\log (1-{\bf{Q}}^k)^{-1}
\end{eqnarray}
with the partition function
\begin{eqnarray}
Z=e^{\sum_{k=1}^\infty \log(1-{\bf{Q}}^k)^{-1}}=\prod_{k=1}^\infty\frac{1}{1-{\bf Q}^k}\, .
\end{eqnarray}


\bibliographystyle{unsrt}
\bibliography{References}

\end{document}